\documentclass[twocolumn]{aastex631}

\usepackage{amsmath}
\usepackage{graphicx}
\usepackage{xcolor}
\usepackage{hyperref}

\renewcommand{\d}[0]{\mathrm{d}}

\shorttitle{Cosmologically coupled compact objects}
\shortauthors{Croker, Zevin, Farrah, Nishimura, and Tarl\'e}

\begin{document}

\title{Cosmologically coupled compact objects: a single parameter model for LIGO--Virgo mass and redshift distributions}

\author[0000-0002-6917-0214]{Kevin.~S.~Croker}
\affiliation{Department of Physics and Astronomy, University of Hawai`i at M\=anoa, 2505 Correa Rd., Honolulu, HI, 96822}

\author[0000-0002-0147-0835]{Michael.~J.~Zevin}
\altaffiliation{NASA Hubble Fellow}
\affiliation{Kavli Institute for Cosmological Physics, The University of Chicago, 5640 South Ellis Avenue, Chicago, IL 60637}
\affiliation{Enrico Fermi Institute, The University of Chicago, 933 East 56th Street, Chicago, IL 60637}

\author[0000-0003-1748-2010]{Duncan~Farrah}
\affiliation{Department of Physics and Astronomy, University of Hawai`i at M\=anoa, 2505 Correa Rd., Honolulu, HI, 96822}
\affiliation{Institute for Astronomy, University of Hawai`i,  2680 Woodlawn Dr., Honolulu, HI, 96822 }

\author[0000-0001-8818-8922]{Kurtis~A.~Nishimura}
\affiliation{Department of Physics and Astronomy, University of Hawai`i at M\=anoa, 2505 Correa Rd., Honolulu, HI, 96822}

\author[0000-0003-1704-0781]{Gregory~Tarl\'e}
\affiliation{Department of Physics, University of Michigan, 450 Church St., Ann Arbor, MI, 48109 }

\begin{abstract}
  We demonstrate a single-parameter route for reproducing higher mass objects as observed in the LIGO--Virgo mass distribution, using only the isolated binary stellar evolution channel.
  This single parameter encodes the cosmological mass growth of compact stellar remnants that exceed the Tolman-Oppenheimer-Volkoff limit.
  Cosmological mass growth appears in known solutions to General Relativity with cosmological boundary conditions.
  We consider the possibility of solutions with cosmological boundary conditions, which reduce to Kerr on timescales short compared to the Hubble time.
  We discuss complementary observational signatures of these solutions that can confirm or invalidate their astrophysical relevance.
\end{abstract}

\keywords{Astrophysical black holes (98), Supernova remnants (1667), Compact objects (288), Cosmological evolution (336), Gravitational wave astronomy (675)}

\section{Introduction}
\label{sec:intro}
\begin{figure*}
\includegraphics[width=\textwidth]{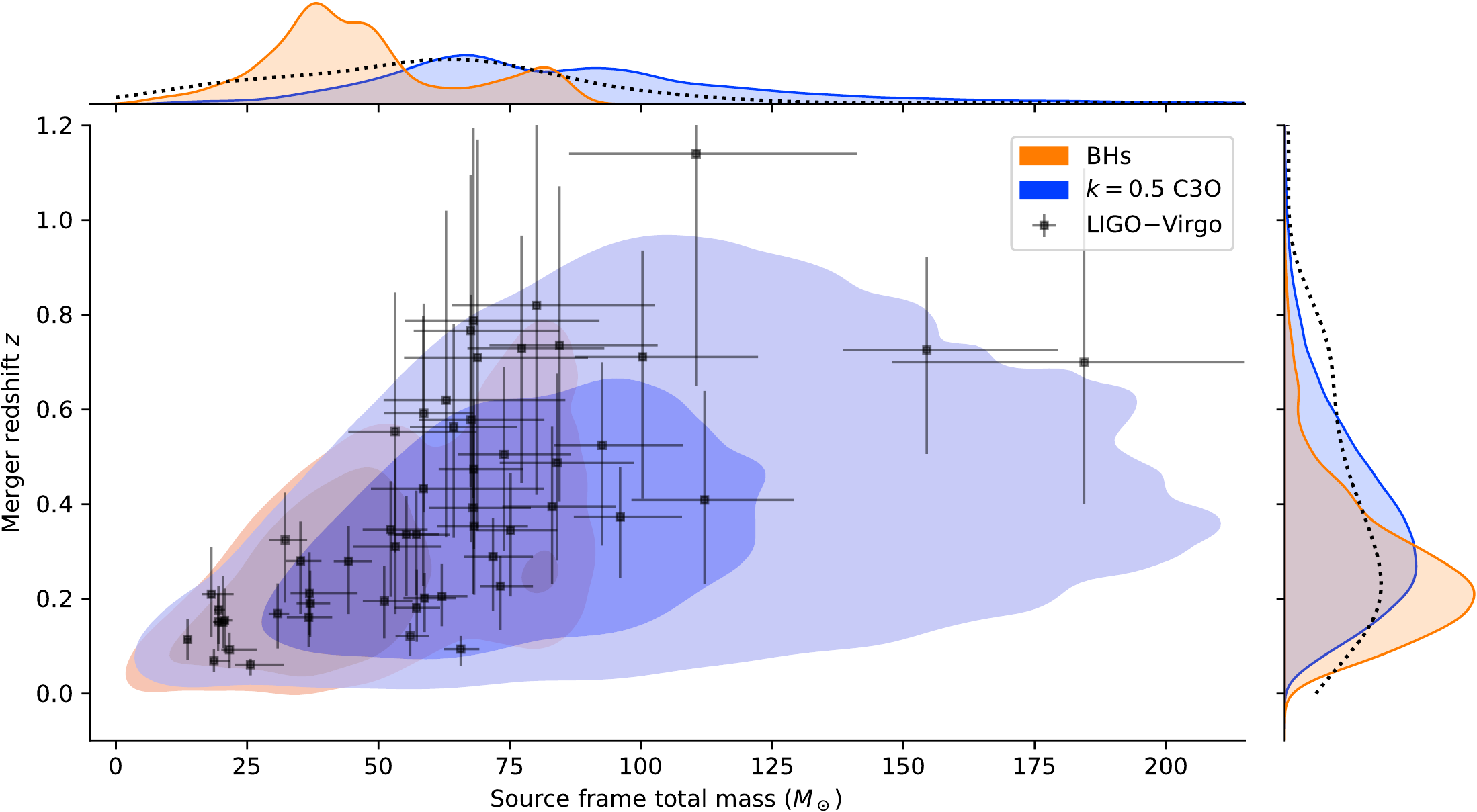}
\caption{\label{fig:k0.5}
  Observable merger redshift and source frame total mass joint distribution of a decoupled stellar remnant population ($k=0$, orange) and the same stellar remnant population evolved as cosmologically coupled objects ($k=0.5$, blue).
  Contours display $1\sigma$ (dark) and $2\sigma$ (light) confidence.
  Predicted detectable populations are computed from the same $10^6$ input stellar remnant binaries.
  The decoupled population produced $8417$ observable mergers, while the $k=0.5$ population produced $167867$, a $20\times$ increase in merger rate.
  The most recent LVC sample of binary mergers (black) is overplotted, with 90\% credible intervals in total mass and redshift shown~\protect{\citep{abbott2019gwtc, abbott2021gwtc, abbott2021gwtc21}}.
  Marginalized distributions in merger redshift and source frame total mass are displayed on the right and top axes, respectively, with the LVC population given for reference (dashed, black).
  LVC binary systems with primary or secondary masses below $2.5M_\odot$ have been excluded.
  Credible intervals at $90\%$ have been truncated for the two extreme LVC systems for clarity of visualization.
 }
\end{figure*}
There is overwhelming evidence for the existence of ultrarelativistic compact objects, both as the end stages of stellar collapse and as the supermassive compact objects in the centers of galaxies.
These compact objects are interpreted within the framework of General Relativity (GR), which has withstood tests on scales ranging from cosmological to the binary pulsar.
A contemporary triumph in this field, following the development of numerical methods for modeling gravitational radiation from compact object mergers \citep{sperhake2015numerical}, was the detection of such mergers by gravitational wave observatories \citep{abbott2016observation}.
The LIGO Scientific Collaboration and the Virgo Collaboration (LVC) has currently published over 50 confident detections of compact binary mergers, the majority being consistent with Kerr black holes (BHs) \citep{abbott2019gwtc, abbott2021gwtc, abbott2021observation}.

The mass spectrum of these objects is less straightforward to interpret.
It was expected \cite[e.g.][]{bethe1998evolution} that isolated binary stellar evolution would be a dominant channel for synthesizing binary black hole mergers visible by LIGO--Virgo.
In this scenario, massive progenitor stars in a binary system proceed through one of more phases of stable and/or unstable mass transfer during their evolution that can significantly tighten the orbit of the resultant black hole binary.
Simulations investigating the population properties of such systems \citep[e.g.][]{dominik2012double} predicted a relatively narrow mass spectrum near $\sim 8M_\odot$ at solar metallicity and a broader spectrum that pushes up to $\sim 35M_\odot$ for subsolar populations. 
Though lower metallicities lead to more massive remnants through weakened wind mass loss, the pair instability process in the cores of massive stars is expected to cause a dearth in the remnant mass spectrum between $\sim 50-120 M_\odot$~\citep[e.g.][]{woosley2017pulsational, farmer2019mind, marchant2019pulsational}.
The remnants observed by LIGO--Virgo have a broad distribution that contaminates the pair instability mass gap, with several remnants consistent with being above $50M_{\odot}$, and a couple with significant support for component masses above $80M_\odot$ \citep[][]{ligobig,abbott2021population}. 

Several explanations have been proposed to maintain consistency between the LIGO--Virgo mass spectrum and Kerr mergers of stellar remnants.
Systematic explanations include altered mass priors \citep{fishbach2020minding, nitz2021gw190521}. 
Pre-explosion mechanisms include
mass loss suppression in low-metallicity ($Z\lesssim10^{-3}$) progenitors \cite[e.g.][]{liubro20,tani20,belc20,spera19,farrell20,kinugawa20,vink20};
collisions between massive stars in dense clusters \cite[e.g.][]{di2020binary,kremer20,renzo20};
different rate parameterization for ${~}^{12}{\rm{C}} + \alpha \to {}^{16}{\rm{O}} + \gamma$ (i.e. the initial alpha process) \cite[e.g.][]{belczynski2010maximum, belczynski2010effect,farm20,costa20};
different overshoot parameter in Population III stars \cite[e.g.][]{umeda20,tanik20};
or outer layer ejection during the common-envelope phase \cite[e.g.][]{kruc16,klem20}. 
Post-explosion mechanisms for populating the high-mass end of the observed black hole mass spectrum include
hierarchical merging in regions of high stellar density \citep[e.g.][]{anton16,frang20,frang20b,gayat20,liulai20,kimball20,anagn20,gond20,tivari20, rodriguez2019black}; 
accretion in dense gas clouds \citep[e.g.][]{natara20,safar20,ricez20};
accretion and/or hierarchical merging in AGN disks \citep[e.g.][]{mcker12,secun19,yang19,tagawa20};
or sequential merging in triple systems \citep{vig20}.
It has been proposed that multiple formation channels drawn from the above may be needed to explain the diversity of systems in the current observational catalog \citep{zevin2021one}, but it is not clear which combinations can reproduce the LIGO--Virgo mass spectrum at both low and high masses. 
There are also proposed explanations using physics beyond the standard model, including:
pre-explosion mass-loss suppression due to core cooling via new light particles \citep{anasta17,diluzio20,croon20,ngvit20};
an additional energy source (to fusion) throughout the progenitor \citep{ziegler20};
a varying gravitational constant \citep{straight2020modified}; and
efficiently accreting primordial BHs \citep{deluca20,wong20,khalo20}.
 
All of the aforementioned studies share a common thread: they have all assumed that the merging objects are Kerr BHs.
This is reasonable, as the merging binary waveforms are consistent with numerical relativity simulations of two Kerr objects.
Like all BH solutions within GR, however, the Kerr solution is provisional: it features horizons, singularities, and unrealistic boundary conditions \cite[e.g.][]{wiltshire2009kerr}.
The asymptotically flat boundary of Kerr, in particular, is incompatible with established observations.
We inhabit a Universe that precisely agrees with the predictions of Robertson-Walker (RW) cosmology at large distances \citep{aghanim2020planck}.
In other words, Kerr can only be consistently interpreted as an approximation to some astrophysically realized solution, appropriate for intervals of time short compared to the Hubble time $t_\mathrm{H}$.

Attempts to construct solutions, which capture such astrophysical aspects, have made some progress.
For example, gravitational collapse to the gravastar solution produces a non-singular, horizon-free, null surface wrapping a stable bubble of Dark Energy (DE) \citep{mazur2015surface, beltracchi2019formation}.
While the gravastar solution does not spin, it has been shown to provide the only known material source to Kerr exterior spacetimes to first \citep{posada2017slowly} and second \citep{beltracchi2021slowly} order in slow-rotation extensions.
With respect to correct boundary conditions, there are many known exact GR solutions describing various objects embedded within cosmologies.
For example, the \citet{mcvittie1933mass} analogue of a Schwarzschild BH, and the matched \citet{nolan1998point} interior solution, exist within an arbitrary RW cosmology.

The presence of cosmological boundary conditions opens new frontiers for dynamics in time.
Locally, solutions with dynamical gravitating mass and horizons that comove with the cosmological expansion have been constructed \citep{faraoni2007cosmological}.
These solutions are significant because they are explicit counter-examples to arguments that local evolution must occur decoupled from cosmological evolution \cite[e.g.][]{einstein1945influence, einstein1946corrections, weinberg2008cosmology, peebles2020large}.
Recent global results in GR are consistent with these findings.
It has been shown that a population of objects, over which the averaged pressure does not vanish, must couple cosmologically \citep{cw2018part1}. 
For example, pure DE objects acquire a dynamical gravitating mass proportional to the RW scale factor $a$, cubed.
Given number densities that diminish $\propto 1/a^3$, such a population then mimics a cosmological constant \citep{crokrunfarr}.
The relativistic effect is entirely analogous to the cosmological photon redshift.

In this paper, we consider the consequences of cosmological coupling within the compact objects observable by LIGO--Virgo.
We restrict our attention to objects in excess of the Tolman-Oppenheimer-Volkoff (TOV) limit, as neutron stars are governed (in principle) by known physics.
We will refer to these objects as Cosmologically Coupled Compact Objects, or C3O for short.
This work extends preliminary investigations of pure DE objects \citep{croker2020implications} and considers generic C3Os in the context of current LIGO--Virgo detection sensitivities. 
We use a contemporary stellar population synthesis code to produce a fiducial compact binary population from the isolated binary evolution channel.
We then track the orbital evolution of each member of the population, determining merger redshift and source-frame masses.
Finally, we incorporate semi-analytic estimates to the sensitive spacetime volume appropriate for the LIGO--Virgo detector network during its recent observational runs and compare the properties of the C3O population with the current observational sample of compact binary mergers.
\section{Model}
\label{sec:model}
We adopt the following single parameter model of cosmological coupling
\begin{align}
  m(a) := m_0\left(\frac{a}{a_i}\right)^k \qquad a \geqslant a_i. \label{eqn:model}
\end{align}
Here $m_0$ is the active gravitational mass of the input stellar remnant, $a$ is the scale factor, $a_i$ is the scale factor at which the input stellar remnant was formed, and $k$ is a dimensionless constant.
This form is motivated by the known cosmological energy shift in photons $(k=-1)$, and the predicted cosmological energy shift in pure Dark Energy objects $(k=3)$.
Note that $k\to 0$ is the decoupled limit, $k < 0$ gives a cosomological energy loss, and $k > 0$ gives a cosmological energy gain.
From global analysis, physically realistic values of $k$ are constrained to
\begin{align}
  -3 \leqslant k \leqslant 3,
\end{align}
by the requirement that all causal observers perceive causal flux.

For each input binary system, we numerically integrate the orbit through the linear radiative regime.
Let $R$ denote semi-major axis, $L$ angular momentum, $e$ eccentricity, $M$ primary mass, and $q$ mass ratio.
The typical linear evolution equations are derived assuming that $M$ does not evolve in time.
We extend these equations to C3O systems.
Because cosmological coupling introduces no preferred spatial directions, angular momentum and eccentricity of the binary are unaffected.\footnote{This can also be shown directly with the theory of adiabatic invariants, see \citet{croker2020implications}.}
This means that $\d L/\d a$ and $\d e/\d a$ are unaltered from the standard expressions of \citet{peters1964gravitational}, apart from substitution of the time-dependent mass.
The remaining evolution equation for semi-major axis then follows from the chain rule,
\begin{align}
  \frac{\d R}{\d a} = \underbrace{\frac{\partial R}{\partial L}\frac{\d L}{\d a}\Bigg|_{e,M} + \frac{\partial R}{\partial e}\frac{\d e}{\d a}\Bigg|_{L,M}}_\text{Radiative decay} + \underbrace{\frac{\partial R}{\partial M}\frac{\d M}{\d a}\Bigg|_{e,L}}_\text{Adiabatic decay}. \label{eqn:evolution}
\end{align}
The radiative decay portion is the standard expression for $\d R/\d a$, substituted with time-dependent mass.
The adiabatic decay portion is computed from conservation of angular momentum in Newtonian binary evolution, combined with the derivative of Equation~(\ref{eqn:model}).

Note that we have omitted explicit reference to $q$ in Equation~(\ref{eqn:evolution}).
In any realistic binary system, one object will become a C3O before the other.
The timescale of this delay, however, is extremely short compared to the Hubble time $t_\mathrm{H}$ for stellar progenitors leading to compact binaries observable by the LIGO--Virgo network.  
We thus neglect this difference and regard both objects as having converted at the same redshift.
This implies that $q$ is non-dynamical.

\section{Methods}
\label{sec:methods}
\begin{figure}[t]
\includegraphics[width=\columnwidth]{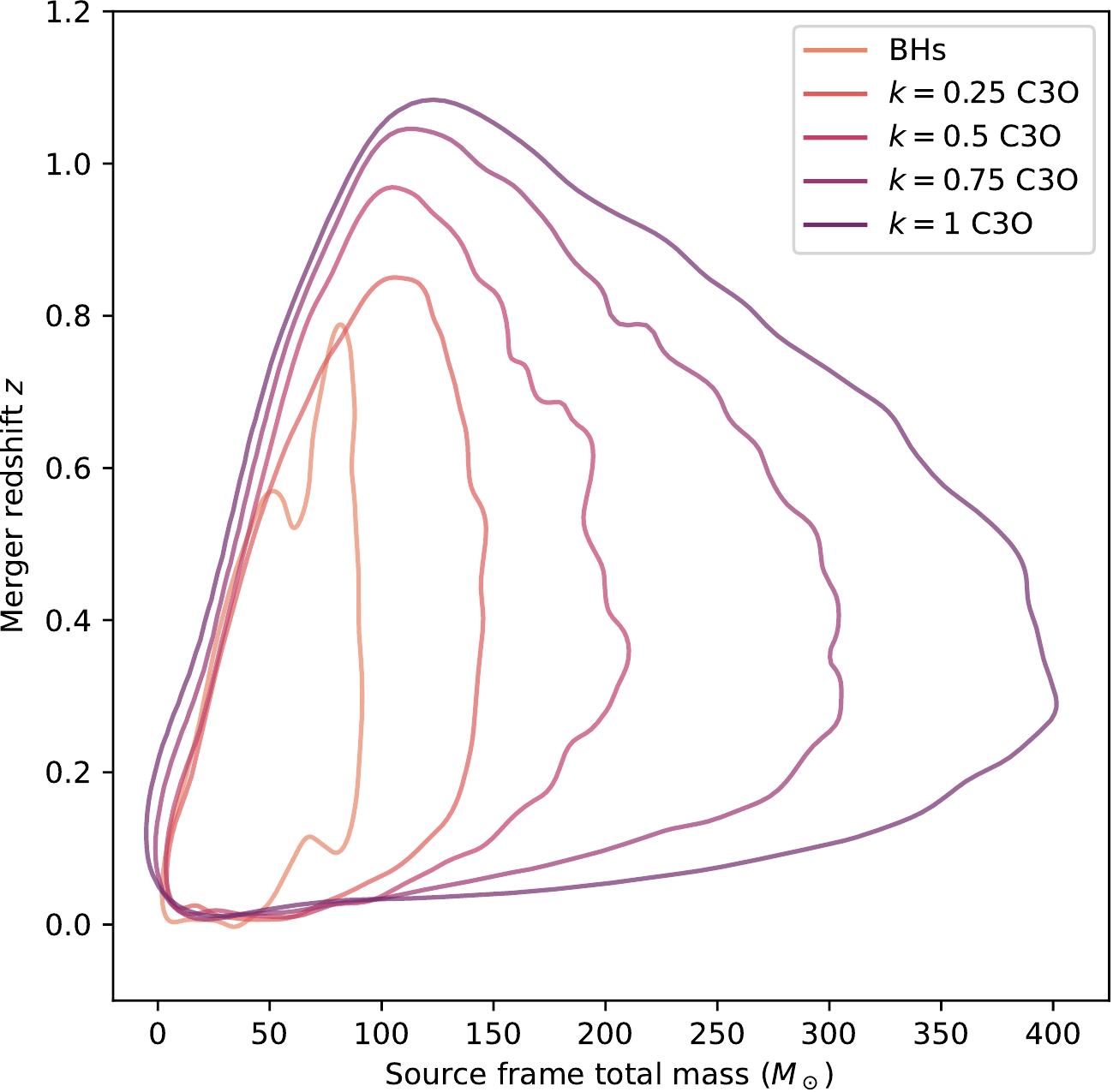}
\caption{\label{fig:variable_k}
  Observable redshift and source frame total mass joint distribution confidence intervals at $2\sigma$ for a grid of cosmological coupling strengths $0 \leqslant k \leqslant 1$.
  Darker color indicates increasing strength in steps of $\Delta k = 0.25$.
  Black holes correspond to $k=0$ and do not couple.
}
\end{figure}%
The presented model requires an input stellar remnant population distributed over formation redshift. 
For simplicity, we assume that the full population of compact remnants originates from the canonical isolated binary evolution channel, which includes hardening of the compact binary progenitor through either stable mass transfer or common envelope phase. 
Though a large number of uncertainties exist in isolated binary evolution that can affect the population properties and merger rates of compact remnants (e.g., winds, mass transfer stability and efficiency, common envelope onset and evolution, natal kicks, remnant mass prescriptions, star formation history and metallicity evolution), we simulate a single model with standard assumptions using the open-source binary
population synthesis code \textsc{COSMIC}\footnote{\url{cosmic-popsynth.github.io}, Version 3.4} \citep{breivik2020cosmic} and focus on the impact of cosmological coupling on this fiducial population. 

Our isolated binary population uses default parameterizations of \textsc{COSMIC} for the majority of its settings \citep[see][and reference therein for details]{breivik2020cosmic}. 
We simulate 16 populations with metallicities equally log-spaced between $Z=0.0001$ and $Z=0.03$. 
As cosmological coupling can cause systems with gravitational-wave inspiral times greater than a Hubble time $t_{\rm H}$ to merge, we determine whether our population is converged using the properties of binary black hole systems that are both merged systems and those that are unmerged after $t_{\rm H}$. 
Critical mass ratios for the onset of unstable mass transfer are implemented following
\citet{neijssel2019effect},
and we assume a ``pessimistic'' common envelope survival scenario~\citep[e.g.][]{dominik2012double}. 
We use a maximum neutron star mass (i.e. TOV limit) of $2.5 M_\odot$ with a delayed remnant mass prescription~\citep{fryer12} as implemented in 
\citet{zevin2020exploring}.
We create a resampled population of $10^6$ binaries from the 16 metallicity runs and populate these systems in redshift using the star formation rate density and mean metallicity redshift dependencies in 
\cite{madau2017radiation}, where we assume a log-normal distribution about the mean metallicity at a given redshift with a dispersion of 0.5 dex. 
We numerically integrate the evolution equations in the C3O scenario until merger, or stop following systems that have not merged by $z=0$.

To compare the output population with the compact binaries observed by LIGO--Virgo, we estimate the relative sensitive spacetime volume of each simulated merger given its masses and merger redshift. 
We pre-compute detection probabilities for a grid of systems in primary mass, mass ratio, and redshift space assuming a 3-detector network consisting of LIGO-Hanford, LIGO-Livingston, and Virgo with \textsc{midhighlatelow} power spectral densities~\citep{abbott2020prospects}.
Using this grid, we utilize a nearest neighbors algorithm to determine the detection probability of each system in our population based on its masses and redshift. 
The relative detection weighting of each system also accounts for the surveyed spacetime volume, such that the relative weight of each system $i$ is 
\begin{equation}
    w_i = p_{\mathrm{det}}\left(M_{i},q_{i},z_{i}\right) \frac{\d V_c}{\d z} \frac{\d t_\mathrm{m}}{\d t_0},
\end{equation}
where $M_{i}$ is the primary mass of system $i$, $q_{i} \leqslant 1$ is the mass ratio, $\d V_c/\d z$ is the differential comoving volume at redshift $z_{i}$, and $\d t_\mathrm{m}/\d t_0 = (1+z)^{-1}$ is the time dilation between clocks at the merger and clocks on Earth. 

\section{Results}
\label{sec:results}
To showcase the utility of the model, we present a C3O population $(k = 0.5)$ and a decoupled $(k = 0)$ population in Figure~\ref{fig:k0.5}.
These observable populations begin with identical input stellar remnant populations.
The known tensions between the LVC population and BHs, as anticipated from the isolated binary evolution channel, are apparent.
In particular, the lack of systems with a source-frame total mass $\gtrsim 90 M_\odot$ is driven by the pair instability process limiting the mass of component BHs to $\lesssim 45 M_\odot$. 
If evolved as C3O, remnants formed through this same channel show good qualitative agreement.
The broad distribution in source frame total mass of C3O is much closer to the LVC population in slope and peak location.
The C3O population also reproduces the LVC population preference for higher redshift mergers.
This is an expected correlation in mass and redshift from observational bias; massive systems are easier to detect, while more systems will be found at the detection horizon where sensitive volume is largest.

Dependence of the observable population on the coupling strength $k$ is shown in Figure~\ref{fig:variable_k}.
The trend is toward a smoother distribution with increased merger redshift and progressively larger masses as the coupling strength increases.
The horizon redshift levels off since massive mergers that would be visible at this horizon are redshifted out of band. 

\section{Discussion}
\label{sec:discussion}
Cosmological coupling in binary systems directly alters their rate of orbital period decay.  
Electromagnetic observation of a putative BH-Pulsar binary over many periods could allow constraint at levels comparable to the Hulse-Taylor system, PSR B1913+16.
This system has been constrained to $\pm 0.2\%$ of the GR prediction after $35$ years of observation \citep{weisberg2010timing}.
For comparison, a binary C3O with $-0.3 \lesssim k \lesssim -0.15$ can lead to $\sim 1\%$ alterations in orbital period decay \citep{cw2018part1}.
The upcoming space-based gravitational wave observatory, LISA, will be able to directly measure the rate of orbital period decay for putative BH-BH binaries in the intermediate mass regime \citep{danzmann2017lisa}.
The sensitivity of LISA to cosmological coupling is the topic of future work.

In this work, we only consider a single astrophysical model of the remnant population in order to succinctly demonstrate the impact of cosmological coupling.
Our input model may, of course, not be an accurate representation of the birth parameters and redshift evolution of binary BH systems. Contributions from other formation scenarios~\citep[e.g.][]{zevin2021one} and uncertainties in the isolated evolution channel itself~\citep[e.g.][]{belczynski2021uncertain}
can strongly affect this input population. 
One of the most qualitatively interesting features of the C3O scenario is its ability to populate the high end of the mass spectrum, where a dearth of BHs is expected in the isolated evolution channel due to the pair instability process.
Both individual events and analysis of the full binary BH population hint at structure beyond a sharp high-mass cutoff in the BH mass spectrum~\citep{abbott2021population}.
Though the presence of a mass gap in the isolated evolution paradigm is a robust prediction, the exact location and width of the gap is uncertain~\citep[e.g.,][]{farmer2019mind}.
Other channels are also predicted to be more efficient at generating merging BHs that occupy this gap, such as stellar mergers or hierarchical BH mergers in dense stellar environments.
Thus, values of the coupling parameter $k$ that match the observational population of merging BHs will be sensitive to the input birth population of binary BHs.
What we have shown is that C3O solutions may act as a facade for the high mass end of the BH birth mass spectrum.
Occurrence of such solutions in nature can ease the tension between the observed BH population and astrophysical models of remnant formation.

The LIGO--Virgo population of merging compact remnants provides an unprecedented window into compact object physics. 
Compact objects with realistic, cosmological, boundary conditions can experience new dynamics in time.
We have shown that a single parameter model of cosmological mass growth, subsequent to compact remnant formation through the isolated binary evolution channel, provides good qualitative agreement with the source frame total masses and merger redshifts of the observed population.
This signature of cosmological coupling can be directly measured from the rate of orbital period decay, and may be accessible to next generation gravitational wave observatories.

\begin{acknowledgments}
  K.~Croker thanks N.~Warrington and the University of Washington Institute for Nuclear Theory (INT) for hospitality during preparation of this work.
  M.~Zevin is supported by NASA through the NASA Hubble Fellowship grant HST-HF2-51474.001-A awarded by the Space Telescope Science Institute, which is operated by the Association of Universities for Research in Astronomy, Inc., for NASA, under contract NAS5-26555. 
\end{acknowledgments}

\bibliography{c3o}{}
\bibliographystyle{aasjournal}
\end{document}